\documentclass[10pt, conference]{IEEEtran}

\usepackage{amsmath}
\usepackage{amsfonts,amssymb}
\usepackage{mathrsfs}
\usepackage{bm}
\usepackage{indentfirst}
\usepackage{graphicx}
\usepackage{epstopdf}
\usepackage[font=small, labelsep=period]{caption}
\usepackage{algorithm, algorithmic}
\usepackage{color}
\usepackage{booktabs}
\usepackage{threeparttable}
\usepackage{cite}
\usepackage{subfig}
\usepackage{array}

\begin{document}

\title{IMNet: A Learning Based Detector for Index Modulation Aided MIMO-OFDM Systems}

\author{\IEEEauthorblockN{Jinxue Liu, Hancheng Lu}
\IEEEauthorblockA{University of Science and Technology of China, Hefei, China, 230027 \\
Email: jxliu18@mail.ustc.edu.cn, hclu@ustc.edu.cn}}

\maketitle

\begin{abstract}
Index modulation (IM) brings the reduction of power consumption and complexity of the transmitter to classical multiple-input multiple-output orthogonal frequency division multiplexing (MIMO-OFDM) systems. However, due to the introduction of IM, the complexity of the detector at receiver is greatly increased. Furthermore, the detector also requires the channel state information at receiver, which leads to high system overhead. To tackle these challenges, in this paper, we introduce deep learning (DL) in designing a non-iterative detector. Specifically, based on the structural sparsity of the transmitted signal in IM aided MIMO-OFDM systems, we first formulate the detection process as a sparse reconstruction problem. Then, a DL based detector called IMNet, which combines two subnets with the traditional least square method, is designed to recover the transmitted signal. To the best of our knowledge, this is the first attempt that designs the DL based detector for IM aided systems. Finally, to verify the adaptability and robustness of IMNet, simulations are carried out with consideration of correlated MIMO channels. The simulation results demonstrate that the proposed IMNet outperforms existing algorithms in terms of bit error rate and computational complexity under various scenarios. \\

\end{abstract}


\section{Introduction}
To meet the high requirements of spectrum efficiency (SE) and energy efficiency for next generation wireless communication systems, index modulation (IM) \cite{IM-Mag} has attracted extensive attention and research in recent years. IM is a general term for a series of innovative modulation technologies, which convey information bits not only on the modulated symbols but also on the on-off status of some resource blocks (such as antennas, subcarriers and time slots) \cite{JSAC-SM,OFDM-IM,SM-Space-Time}. Due to that some of resource blocks are inactive (off), IM can bring the reduction of energy consumption and complexity of the transmitter at the expense of acceptable degradation of SE.\par

In view of above characteristics, IM has been widely applied to existing wireless communication systems in different ways. Among them, generalized spatial modulation (GSM) \cite{GSM} and orthogonal frequency division multiplexing with IM (OFDM-IM) \cite{OFDM-IM} are the two typical cases in the IM family, which transfer part of the information bits on the indices of the active transmit antennas (TAs) and subcarriers, respectively. To further improve SE, \cite{MIMO-OFDM-IM} combines OFDM-IM with multiple-input multiple-output (MIMO) systems, while \cite{Space-Time-TVT} and \cite{Two-Dim-TVT} apply IM across multiple domains (such as space, frequency and time domains). Compared with classical MIMO-OFDM systems, the above IM aided systems can improve the  bit error rate (BER) performance. Meanwhile, it can reduce inter-channel interference, peak to average power ratio and power consumption and relax inter-antenna synchronization requirements.\par

It is worth noting that the traditional detectors in classical MIMO-OFDM systems cannot be directly applied to the IM aided MIMO-OFDM (IM-MIMO-OFDM) systems. Since part of the information bits are conveyed on the indices of the acitve TAs and subcarriers, not only the modulated symbols but also the indices of the acitve TAs and subcarriers need to be detected, which makes the complexity of detector greatly increased. \cite{SM-ML} and \cite{OFDM-IM-ML} have analyzed the maximum likelihood based detector (MLD), but the complexity of the detector increases exponentially with increasements of the numbers of TAs and subcarriers and the signal constellation size. To further reduce the detection complexity, several low complexity detectors have been proposed, such as, simple minimum mean square error (MMSE) detector, matched filtering (MF) detector, log-likelihood ratio (LLR) detector and signal vector based list (SVBL) detector. However, compared to the MLD, these low complexity detectors suffer from a significant error performance degradation. Particularly, all the aforementioned methods are heavily dependent on the channel state information at receiver (CSIR), which requires high system overhead in the practical communication scenarios.\par

To be practical, imperfect CSIR should be considered in the design of detectors. For carrying out the detection at receiver with imperfect CSIR, deep learning (DL), which has the powerful feature extraction and generalization abilities, can be introduced in the detector design. With these abilities, DL based detectors can realize robust detection in MIMO systems\cite{MIMO-MOD-DL,Learn-to-detect}. Apart from this, compared with the traditional detectors, DL based detectors are non-iterative, which show great computational complexity reduction\cite{ComNet}.\par


\begin{figure*}[htbp]
	\centering
	\includegraphics[width=1\textwidth]{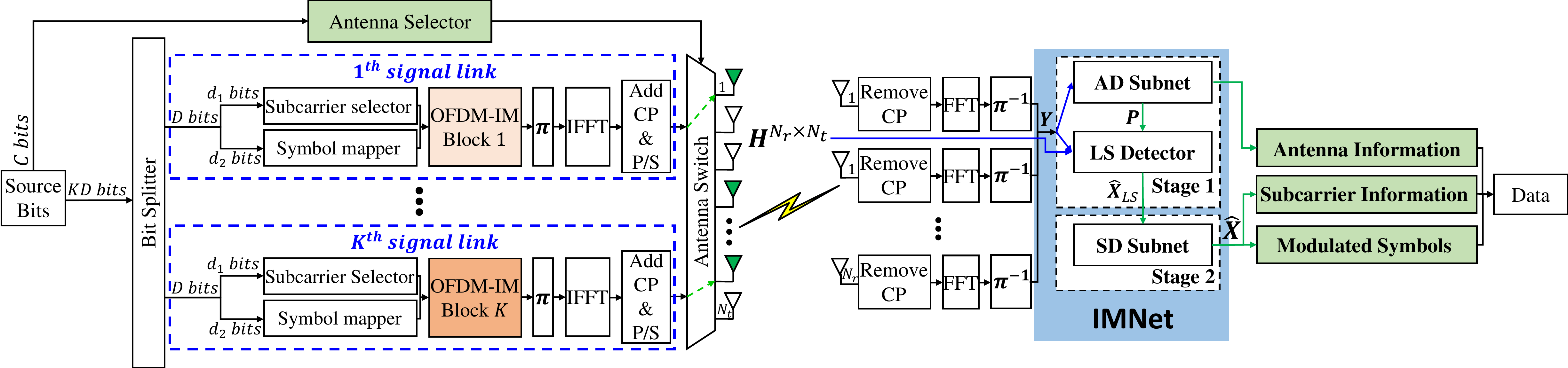}
	\caption{The block diagram of transmitter and DL based receiver in IM-MIMO-OFDM systems}
	\label{fig-2.1}
\end{figure*}

In this paper, we consider the low complexity and robust detector design for the IM-MIMO-OFDM systems. The information bits in the IM-MIMO-OFDM systems are conveyed on three parts: the indices of the active TAs, the indices of the active subcarriers and the modulated symbols, which are also the three parts that need to be detected at receiver. In this case, the traditional detectors face high complexity. However, the transmitted signal in IM-MIMO-OFDM systems actually has the property of structural sparsity. Based on this property, the detection process at receiver is formulated as a sparse reconstruction problem. The formulated sparse reconstruction problem can be solved through the following two steps. The first step is estimating the indices of non-zero elements in the transmitted signal vector through the received signal vector. The second step is estimating the values of the non-zero elements with the prior knowledge of their indices. Considering the imperfect CSIR and computational complexity, a DL based non-iterative detector called IMNet is proposed to realize the above sparse reconstruction process. The proposed IMNet consists of two subnets (i.e., antennna detection (AD) subnet and signal denoising (SD) subnet) as well as a least square (LS) detector. The task of the AD subnet is to realize the first step of the sparse reconstruction process, and the second step is accomplished through combining the AD subnet with the LS detector, which callded as DL based matching pursuit (DLBMP). For further improving the performance of the IMNet, the SD subnet is introduced to remove the noise effects. \par
To verify the adaptability and robustness of the proposed IMNet, we perform simulations with consideration of two channel models (i.e., the Rayleigh fading MIMO channel and the correlated MIMO channel) and two CSIR conditions (i.e., the perfect CSIR and the imperfect CSIR). The simulation results demonstrate that the IMNet has a better BER performance and robustness than the traditional algorithms in different scenarios, and the computational complexity is much smaller than the traditional algorithms. \par
The rest of this paper is organized as follows. The system model and problem formulation are given in Section \ref{sec-2}. Section \ref{sec-3} details the architecture of the proposed IMNet. The simulation results are presented and discussed in Section \ref{sec-4}. Finally, conclusions are drawn in Section \ref{sec-5}.

\section{System Model and Problem Formulation}\label{sec-2}
In this section, we first demonstrate the system model of the IM-MIMO-OFDM systems in Section \ref{sec-2.1}. Then, the sparse reconstruction problem is formulated in Section \ref{sec-2.2}.

\subsection{System Model}\label{sec-2.1}
We consider an IM-MIMO-OFDM system that equipped with $ N_{t} $ TAs, $ N_{r} $ receive antennas (RAs) and $ N_f $ subcarriers, whose block diagram is depicted in Fig.\ref{fig-2.1}. At transmitter, IM is applied to both space and frequency domains, which means the information bits are conveyed not only over the modulated symbols but also over the indices of the active TAs and subcarriers. At receiver, these transmitted information bits can be recovered through the proposed IMNet.\par

As shown in Fig.\ref{fig-2.1}, each IM-MIMO-OFDM frame is comprised of a total number of $ C+KD $ incoming data bits. The first $ C = \lfloor \log_2 \binom{N_t}{K} \rfloor$ bits are used to select $ K $ active antennas from $ N_{t} $ TAs for transmitting.\footnote{$\binom{N}{K}$ and $ \lfloor\cdot\rfloor $ denotes the binomial coefficient and floor function, respectively. The mapping between the $ C $ bits and the TA combination patterns can be implemented by using a look-up table \cite{GSM}.} In this way, $ K $ signal links will be generated while the remaining $ N_t - K $ TAs keep off status. The left $ KD $ bits are equally split into $ K $ blocks and separately processed on the $ K $ signal links. Unlike classical OFDM systems, which map all data bits to constellation points for all subcarriers, the $ D $ bits on each signal link are divided into two parts. The first part with $ d_1 = \lfloor \log_2 \binom{F}{N_f} \rfloor $ bits is used to select $ F $ active subcarriers from all the $ N_f $ subcarriers to convey information, while the remaining $ N_f-F $ subcarriers are set to be idle.\footnote{The mapping between the $ d_1 $ bits and the subcarrier combination patterns can be implemented by using a look-up table or the combinatorial method \cite{OFDM-IM}.} The second part with $ d_2 = F\log_2M $ bits is used to select $ F $ symbols from the signal constellation $ \mathcal{S} $ with $ \left| \mathcal{S} \right| = M $ for $ F $ active subarriers. And then, an OFDM-IM block is generated. After interleaving, inverse fast Fourier transforming (IFFT), adding cyclic prefix (CP) and parallel to serial converting, the $ K $ OFDM-IM blocks are transmitted through $ K $ active TAs. \par
After passing through the MIMO channel, the CP is removed, fast Fourier transform (FFT) and deinterleaving are done at each RA to obtain the received signal in frequency domain. The received signal at the $ i $-th subcarrier for all RAs can be represented as 
\begin{equation}\label{formula-2.1}
\bm{y}_i=\bm{H}_i\bm{x}_i+\bm{w}_i,
\end{equation}

\noindent where $ \bm{y}_i\in\mathbb{C}^{N_r\times1} $ denotes the  received signal vector at the $ i$-th subcarrier, $ \bm{x}_i\in\mathbb{C}^{N_t\times1} $ denotes transmitted signal vector at the $ i $-th subcarrier, $ \bm{H}_i\in\mathbb{C}^{N_r \times N_t} $ denotes the channel frequency response (CFR) between TAs and RAs at the $ i $-th subcarrier and $ \bm{w}_i\in\mathbb{C}^{N_r\times1} $ represents the additive white Gaussian noise (AWGN) vector with zero mean and unit variance at the $ i $-th subcarrier. 

Let $ \bm{\phi}_a $, $ \bm{\phi}_f $ and $ \bm{X}_s $ respectively denote the indices set of the active TAs, the indices set of the active subcarriers and the transmitted signal matrix, where $ \bm{X}_s = \left[ \bm{x}_1, \bm{x}_2, \cdots, \bm{x}_{N_f} \right] $. The MLD\cite{SM-ML} based on Eq.(\ref{formula-2.1}) can be formulated as
\begin{equation}\label{formula-2.2}
	<\bm{\phi}_a, \bm{\phi}_f, \bm{\hat{X}}_s > = \mathop{\arg\min}_{\bm{\phi}_a, \bm{\phi}_f, \bm{X}_s} \sum\limits_{i=1}^{N_f}{\|\bm{y}_i-\bm{H}_i\bm{x}_i\|}^2.
\end{equation}

It should be noted that the MLD detector entails an exhaustive search space based on Eq.(\ref{formula-2.2}). Therefore, there will be a prohibitive complexity for this method because of the large number of transmit antennas, the large size of subcarrier and the complex signal constellation of practical MIMO-OFDM systems.


\subsection{Problem Formulation}\label{sec-2.2}
\begin{figure}[htbp]
	\centering
	\includegraphics[width=0.98\linewidth]{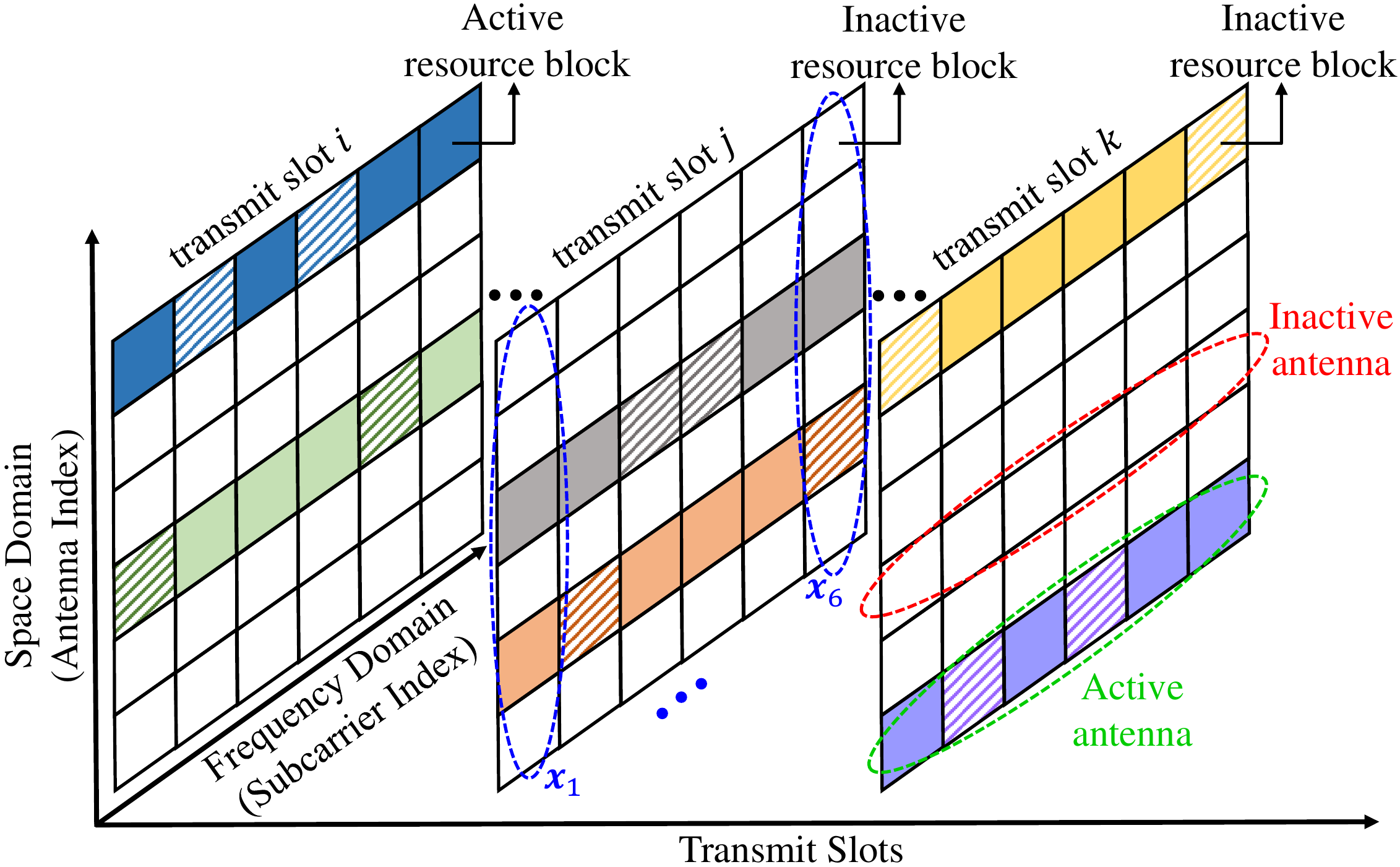}
	\caption{The signal structure of IM-MIMO-OFDM systems in space, frequency and time domains with $ N_t = 6, K=2, N_{f}=6, F=4 $. Colorless squares denote the inactive resource blocks in the space domain, which means these antennas are inactive. The squares with crosshatch in the active space domain represent the inactive resource blocks in the frequency domain, which means such subcarriers are inactive. The squares with pure color denote the active resource blocks that truly convey the modulated symbols. }
	\label{fig-2.2}
\end{figure}

Due to that only $ K $ TAs and $ F $ subcarriers are active for transmitting, the transmitted signal vector $ \bm{x}_i $ is sparse for all subcarriers, which is shown as Fig.\ref{fig-2.2}. By utilizing the sparsity of $ \bm{x}_i $, the detection process can be considered as a sparse reconstruction problem.\par

In this sparse reconstruction problem, $ \bm{x}_i $ is the sparse vector that needs to be recovered, $ \bm{y}_i $ is the measurement vector and $ \bm{H}_i $ is the measurement matrix. Based on Eq.(\ref{formula-2.1}), the sparse reconstruction of $\mathbf{x}_i$ can be formulated as the following optimization problem with the consideration of AWGN.
\begin{equation}\label{formula-2.3}
\hat{\bm{x}_i} = \mathop{\arg\min} {\| \bm{x}_i\|}_1 \ \ s.t. \ {\|\bm{y}_i-\bm{H}_i\bm{x}_i\|}_2 < \varepsilon,
\end{equation}

\noindent where $ \varepsilon $ is a predetermined noise level of the system. There exist many algorithms that can solve such a problem, but they need the measurement matrix $ \bm{H}_i  $ to realize the sparse reconstruction, which cannot be accurately known in practical wireless communication scenarios. In addition, by solving the problem in E.q.(\ref{formula-2.3}), only one transmitted signal vector can be reconstructed at a time, which is inefficient.

To further improve the performance and reduce complexity, we assume that the inactive subcarriers on each active signal link convey a special symbol $ x^{\ast} $, which is shown as squares with crosshatch in Fig.\ref{fig-2.2}. After this operation, the transmitted signal vectors $ \bm{x}_i(1\leq i \leq N_f) $ have the following relationship:
\begin{align}
	&supp(\bm{x}_1)=supp(\bm{x}_2)=\cdots=supp(\bm{x}_{n_f})\label{formula-2.4}, \\
	&{\|supp(\bm{x}_1) \|}_1 = \cdots = {\| supp(\bm{x}_{N_f})\|}_1 = K\label{formula-2.5}.
\end{align}

\noindent Eq.(\ref{formula-2.4}) means all the transmitted signal vectors share the same support\footnote{The support of a vector is the indices of nonzero elements, denoted as $ supp(\cdot) $. For example, if $ \bm{x} = \left[1,0,0,2,0\right] $, $ supp(\bm{x})=\left[1,4\right] $}. Meanwhile, due to the fact that there are $ K $ active antennas, all the transmitted signal vectors are $ K $-sparse, which implies that there are $ K $ non-zero elements (including $ x^\ast $) in each transmitted signal vector, which is expressed as Eq.(\ref{formula-2.5}). Based on these two observations, the signal matrix $ \bm{X}_s $ owns the property of structural sparsity. \par 

Based on the structural sparsity of $ \bm{X}_s $, the detection process of IM-MIMO-OFDM systems can be further formulated as a sparse reconstruction from multiple measurement vectors (MMV). For $\mathbf{X}_s$, the MMV refers to the multiple received signal vectors, i.e., $ \bm{y}_i (1\leq i \leq N_f) $, which is denoted as $ \bm{Y}_s=\left[ \bm{y}_1, \bm{y}_2, \cdots, \bm{y}_{N_f} \right] $. The sparse reconstruction of $ \mathbf{X}_s $ can be formulated as the following optimization problem with the consideration of AWGN.
\begin{subequations}\label{formula-2.6}
	\begin{align}
		\bm{\hat{X}}_s &= \mathop{\arg\min}_{\bm{X}} \ {\parallel supp(\bm{X}) \parallel}_1 \tag{\ref{formula-2.6}}\\
		s.t.\  &Eq.(\ref{formula-2.4}), Eq.(\ref{formula-2.5})\\
		&\sum_{i=1}^{N_f}{\|\bm{y}_i-\bm{H}_i\bm{x}_i\|}_2 < \epsilon
	\end{align}
\end{subequations}
where $supp(\bm{X}) = \bigcup_isupp(\bm{x}_i)$.

By solving this problem in Eq.(\ref{formula-2.6}), the transmitted signal vectors at all subcarriers can be jointly recovered efficiently, compared to separately solve the problem in E.q(\ref{formula-2.3}). The problem can be solved by two steps. The first step is estimating the indices of non-zero rows in $ \bm{X}_s $ through $ \bm{Y}_s $, which also means detecting the indices of active antennas. The second step is estimating the values of the elements in non-zero rows with the prior knowledge of the indices gotten in the first step. The  purpose of the second step is jointly estimating the indices of the active subcarriers and the modulated symbols.\par

Considering the practical situation that $ \bm{H}_i $ cannot be accurately known at receiver and traditional algorithms are iterative with high comolexity, we introduce DL to realize the two steps with imperfect CSIR and finally recover the transmitted signal, which will be detailed in the next section.

\section{Proposed IMNet}\label{sec-3}

In this section, a DL based detector, called IMNet, is designed to recover the transmitted signal of IM-MIMO-OFDM systems, which can significantly decrease the complexity of receiver. The architecture of the IMNet is elaborated in Section \ref{sec3-1}. The details of the AD subnet and combination of the AD subnet and the LS detector are shown in Section.\ref{sec3-2}. The SD subnet is introduced in Section.\ref{sec3-3}. Section.\ref{sec3-4} gives the training specification.

\subsection{IMNet Architecture}\label{sec3-1}
As shown in Fig.\ref{fig-2.1}, the IMNet consists of two convolution neural networks (CNN), which are called as antenna detection (AD) subnet and signal denoising (SD) subnet, respectively. The task of the AD subnet is to predict the activation probabiliy of each TA, while the task of SD subnet is to further improve the precision of the signal that is recovered by LS detector. The working procedure of the IMNet can be divided into two stages. In the first stage, the AD subnet predicts the activation probability of each TA for getting the indices of acitvated TAs, and then, based on the obtained indices and CSIR, the LS detector is used to get the initial estimation of the transmitted signal. In the second stage, in order to refine the initial estimation, the SD subnet is introduced to treat the initial estimation as an image, and remove the noise effects.\par
After above two stages, the indices of activated TAs are obtained and the transmitted signal through each activated TA is recovered. The subcarriers are sorted in descending order according to the amplitude of the recovered symbols on them,  and the first $ F $ subcarriers are considered to be active. Finally, with the indices of the active antennas and subcarriers and modulated symbols, source binary bits can be recovered through inverse mapping.


\begin{figure}[htbp]
	\centering
	\includegraphics[width=0.98\linewidth]{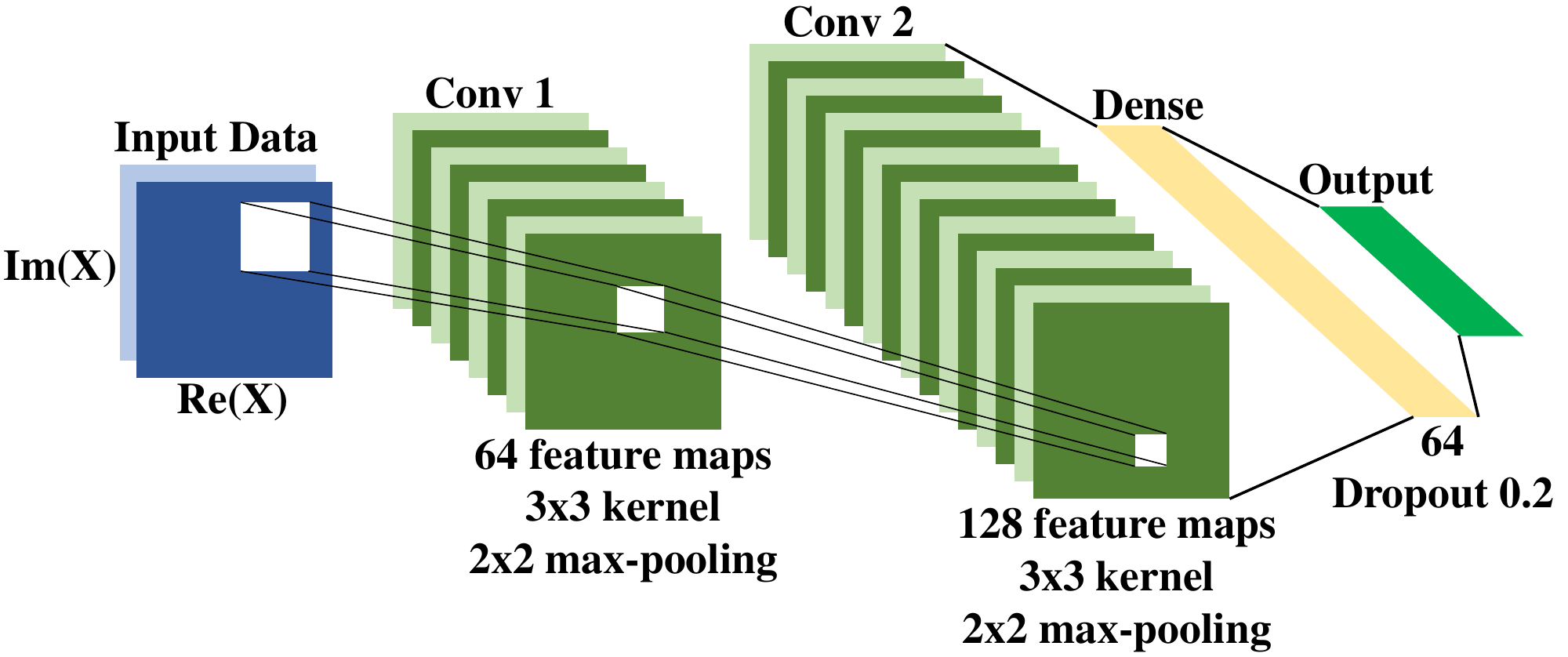}
	\caption{The architecture of AD subnet.}
	\label{fig-3.2}
\end{figure}

\subsection{AD Subnet}\label{sec3-2}

The AD subnet is a four-layer CNN, whose architecture is shown in Fig.\ref{fig-3.2}. For the convolution layers, we use convolution kernels with $ 3\times 3 $ size and Max-Pooling operation with $ 2 \times 2 $ size. The only difference between the two convolution layers is the number of filters. There are $ 64 $ and $ 128 $ filters in the first and second convolution layers, respectively. To accelerate the converging procedure, we apply the Rectified Linear Unit (ReLU) as activation function in the two convolution layers. The general activation function of ReLU is defined as 
\begin{equation}
	ReLU(x_k) = max(x_k, 0),
\end{equation}
\noindent where $ x_k $ is the input signal of the activation on the $ k $-th channel.\par
The input complex valued signal is separated into real and imaginary parts, and then, the two parts are fed into the network, as shown in Fig.\ref{fig-3.2}. The output of the AD subnet is the activation probability of each TA. Therefore, we use the Sigmoid function as the activation function of the output layer and the binary cross-entropy as the loss function, which are expressed in Eq.(\ref{formula-3.2}) and Eq.(\ref{formula-3.3}), respectively.\par
\begin{align}
	&Sigmoid(x_k) = \frac{1}{1+e^{x_k}}\label{formula-3.2}, \\
	L_{AD}(\bm{\theta}_1) = \frac{1}{N_t}&\sum\limits_{i=1}^{N_t}\left[ y_i\log(\hat{y}_i) + (1-y_i)\log(1-\hat{y}_i) \right]\label{formula-3.3},
\end{align}
\noindent where $ \bm{\theta}_1 $ denotes the weights for the AD subnet, $ \hat{y}_i $ and $ y_i $ are the predicted activation probability and the initial on-off label of the $ i $-th TA, respectively.

Compared to the traditional algorithms, the AD subnet can predict the activation probability of each TA and then get the indices of active TAs without the knowledge of CSIR, which produces the excellent performance under various channel conditions.

According to the estimated indices of the activated TAs through the AD subnet, we select corresponding columns in $ {\bm{H}_i} $ to form a new channel matrix $ {\bm{\hat{H}}_i} $. Then, the LS detector can be applied to get the initial estimated signal $ \bm{\hat{x}}_i $, which can be expressed as
\begin{equation}\label{formula-3.4}
\bm{\hat{x}}_i = {\left[ {(\bm{\hat{H}}_i)}^H{\bm{\hat{H}}_i} \right]}^{-1}{(\bm{\hat{H}}_i)}^H{\bm{y}_i},
\end{equation}

\noindent where $ {\left[ {(\bm{\hat{H}}_i)}^H{\bm{\hat{H}}_i} \right]}^{-1}{(\bm{\hat{H}}_i)}^H $ is the pseudo inverse of $ \bm{\hat{H}}_i $. The whole process of the stage 1 which consists of AD subnet and LS can be summarized as Algorithm \ref{alg-1}. $ \tau $ is a predefined threshold to decide whether the TA is active or inactive.

\begin{algorithm}
	\renewcommand{\algorithmicrequire}{\textbf{Input:}}
	\renewcommand{\algorithmicensure}{\textbf{Output:}}
	\caption{Deep Learning Based Matching Pursuit (DLBMP)}
	\label{alg-1}
	\begin{algorithmic}[1]
		\REQUIRE $ \bm{Y} $: the received signal matrix; $ \bm{H} $: CIRF; $ K $: the number of active TAs; $ N_f $: the number of subcarriers.
		\ENSURE $ \bm{\phi}_a $: the indices set of active TAs; $ \hat{\bm{X}}_{LS} $: the recovered signal matrix using LS detector  
		\STATE Initialization: $ \bm{\phi}_a = \varnothing$, $ \hat{\bm{X}}_{LS} = \mathbf{0}^{K \times N_f} $
		\STATE Predict the activation probability of TAs using $ \bm{P} = \left[ p_1, p_2, \cdots, p_{N_t} \right]  = AD(\bm{Y})$
		\STATE Get the indices set of the active TAs: \\ $ \bm{\phi}_a = \left\lbrace i \mid p_i > \tau, 1\leq i \leq N_t \right\rbrace $
		\STATE \textbf{for} $ i=1 \ to \  N_f $ \textbf{do}
		\STATE \ \ According to $ \bm{\phi}_a $, choose columns of $ {\bm{H}_{i}} $ to generate \\ \ \ $ {\bm{\hat{H}}_{i}} $: $ {\bm{\hat{H}}_{i}} =  {\bm{H}_{i}}(\bm{\phi}_i)$
		\STATE \ \ Estimate the signal vector $ \bm{\hat{x}}_{i} $ using Eq.(\ref{formula-3.4})
		\STATE \textbf{end for}
		\STATE $ \hat{\bm{X}}_{LS} = \left[ \bm{\hat{x}}_{1}, \bm{\hat{x}}_{2}, \cdots, \bm{\hat{x}}_{N_f} \right] $
		\RETURN $ \bm{\phi}_a $, $ \hat{\bm{X}}_{LS} $
	\end{algorithmic}
	
\end{algorithm}

\subsection{SD Subnet}\label{sec3-3}
The initial estimation of the transmitted signal is obtained through Algorithm \ref{alg-1}. Considering the CSIR which cannot be accurately known in practical wireless communication scenarios, the initial estimation is coarse. In order to refine the coarse estimation, a denoising network is introduced to mitigate the effects of noise. We choose the state-of-the-art denoising network \cite{DnCNN} in image processing fields as our SD subnet. The input of the SD subnet is the initially estimated signal matrix through Algorithm \ref{alg-1}, and the output is the denoising signal matrix. We adopt the Mean Square Error (MSE) as the loss function for SD subnet. The loss function can be expressed as
\begin{equation}\label{formula-3.5}
	L_{SD}(\bm{\theta}_2)=\frac{1}{T}\sum\limits_{i=1}^{T}\parallel F((\bm{\hat{X}}_{LS})_i; \bm{\theta_2})-\bm{X}_i \parallel^2,
\end{equation}
\noindent where $ \bm{\theta_2} $ denotes the weights of the SD subnet, $ T $ is the number of training pairs, and $ \bm{X}_i $ is the transmitted signal without noise.\par

\subsection{Training The IMNet}\label{sec3-4}

Let $ \bm{\Theta} = \{ {\bm{\theta}}_1,  {\bm{\theta}}_2 \} $ denote the set of all weights of the IMNet. The whole process of the IMNet can be expressed as\par

\vspace{-0.5cm}
\begin{equation}
	\bm{\hat{X}}^i_s = F(\bm{Y}^i;\bm{\Theta}) = F_2(LS(F_1(\bm{Y}^i;\bm{{\theta}}_1),\bm{Y}^i,\bm{H}^i);\bm{{\theta}}_2),
\end{equation}

\noindent where $ F $, $ F_1 $, $ F_2 $ and $ LS $ are the functions of the IMNet, AD subnet, SD subnet and the LS detector, respectively.\par

We use a two stage training algorithm to get the optimal $ \bm{\Theta} $. In the first stage, we train the AD subnet. In the second stage, we freeze the parameters of AD subnet, and train the SD subnet to get the final output. The entire training process is summarized as algorithm \ref{alg-2}.

\begin{algorithm}
	\renewcommand{\algorithmicrequire}{\textbf{Input:}}
	\renewcommand{\algorithmicensure}{\textbf{Output:}}
	\caption{Training the proposed IMNet}
	\label{alg-2}
	\begin{algorithmic}[1]
		\REQUIRE $ \{ \bm{X}^i_s \}^{N}_{i=1} $: dataset of the transmitted signal matrix; $ \{ \bm{Y}_i \}^{N}_{i=1} $: dataset of the received signal matrix; $ \{ \bm{X}^i_a \}^{N}_{i=1} $: labels of the activation status of TAs
		\ENSURE $ \bm{\theta}_1, \bm{\theta}_2 $: Optimal weights 
		\STATE Initialize $ {\theta}_1 $ and $ {\theta}_2 $
		\STATE Train the AD subnet using dateset $ \{ \bm{Y}^i, \bm{X}^i_a \}^N_{i=1} $ to minimize E.q(\ref{formula-3.3}), then freeze $ \bm{\theta}_1 $
		\STATE Use Algorithm \ref{alg-1} to get the initial estimated signal matrix $ \{ \bm{\hat{X}}^i_{LS} \}^N_{i=1} $
		\STATE Train  the SD subnet using dataset $ \{ \bm{X}^i_{LS}, \bm{X}^i_s \}^N_{i=1} $ to minimize E.q(\ref{formula-3.5}), then freeze $ {\theta}_2 $
		\RETURN $ {\theta}_1 $, $ {\theta}_2 $
	\end{algorithmic}	
\end{algorithm}




\section{Performance Evaluation}\label{sec-4}

In this section, we evaluate the performance of IMNet. Simulations are performed to show the superior performance of the proposed IMNet, compared with existing  detection algorithms under various channel conditions. Implementation is first presented, then results are given and discussed. \par

\subsection{Implementation Details}
The two subnets in IMNet are developed through Keras. For the dataset, we obtain the training data through simulations. The two subnets are trained using the stochastic gradient descent method and Adam optimizer on the Nvidia GTX 1080ti GPU environment. The learning rate and batch size are set to 0.001 and 50, respectively.\par 
To verify the adaptability of IMNet, both Rayleigh fading MIMO channel and correlated MIMO channel are considered in our simulations:
\begin{itemize}
	\item \textbf{Rayleigh fading MIMO channel:} The channle matrix $ \bm{H} $ obeys complex Gaussian distribution, i.e., $ \bm{H} \sim \mathcal{CN}(0,\frac{1}{N_t}) $.
	\item \textbf{Correlated MIMO channel:} The channle matrix $ \bm{H} $ can be expressed as
	\begin{equation}
	\bm{H} = \bm\Theta^{1/2}_{Rx}\bm{A}_{iid}\bm\Theta^{1/2}_{Tx}
	\end{equation}
	where $ \bm{A}_{iid} $ is the independent identical distributed (i.i.d.) Rayleigh fading channel, $ \bm\Theta_{Rx} $ and $ \bm\Theta_{Tx} $ are the spatial correlation matrix of TAs and RAs, respectively. In our simulations, the correlation coefficient $ \rho $ is set to 0.5.
\end{itemize}
\par
In order to demonstrate the robustness of the IMNet, imperfect CSIR is also considered in our simulations. Considering the ML channel estimation \cite{Imperfect-CSIR}\cite{CE-ML}, the imperfect CSIR can be expressed as 
\begin{equation}
\bm{\hat{H}} = \bm{H} + \Delta\bm{H},
\end{equation} 

\noindent where $ \bm{\hat{H}} $ and $ \Delta\bm{H} $ denote the imperfect CSIR and estimation error, respectively. $ \Delta\bm{H} $ obeys zero mean i.i.d. complex Gaussian distribution with $ \mathbb{E}\left[ {|\Delta{h}_{m,n}|}^2 \right] = \frac{N_t{\sigma}^2_z}{{N_p}{E_p}} $ \cite{Imperfect-CSIR}, where $ N_p $ and $ E_p$ represent the number and the power of pilot symbols, respectively.\par
We compare IMNet with other two traditional algorithms:\par

(1) ML \cite{SM-ML}\cite{OFDM-IM-ML}: An exhaustive search algorithm that jointly estimates the active antennas, active subcarriers and modulated symbols. \par
(2) Matched Filtering with Log-likelihood Ratio (MF-LLR) : An algorithm that combines MF in \cite{SM} with LLR detector in \cite{wen2017index}. In MF-LLR, the active TAs, the active subcarriers and the modulated symbols are estimated sequentially.\par
\begin{figure}[htbp]
	\centering
	\includegraphics[width=0.8\linewidth]{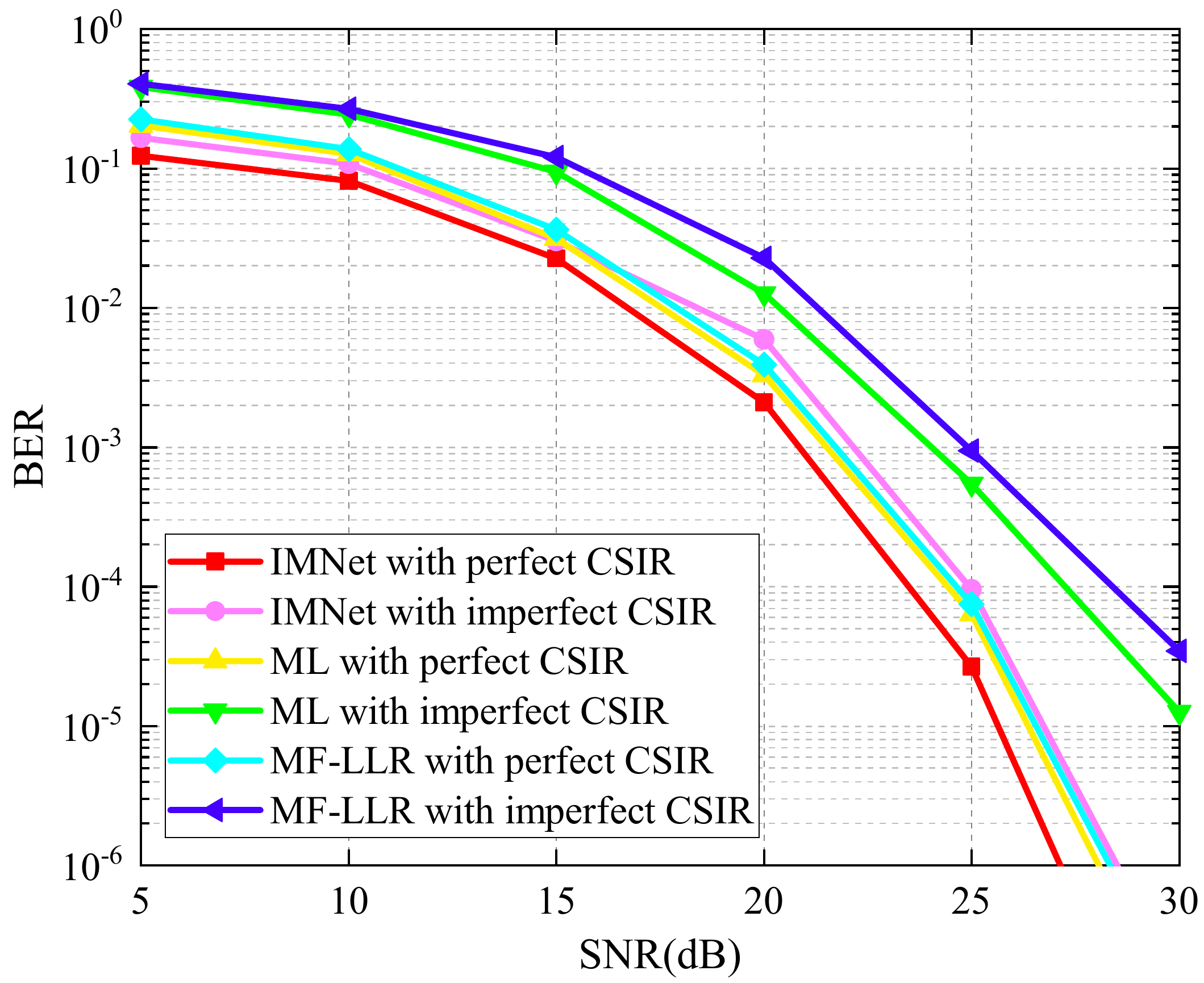}
	\caption{BER performance comparisons of different algorithms with Rayleigh fading MIMO channel. IM-MIMO-OFDM is configured with $ N_t = N_r = 8, N_f = 8, K = 2, F = 6 $ and 4QAM modulation.}
	\label{fig-4.2}
\end{figure}

\begin{figure}[htbp]
	\centering
	\includegraphics[width=0.8\linewidth]{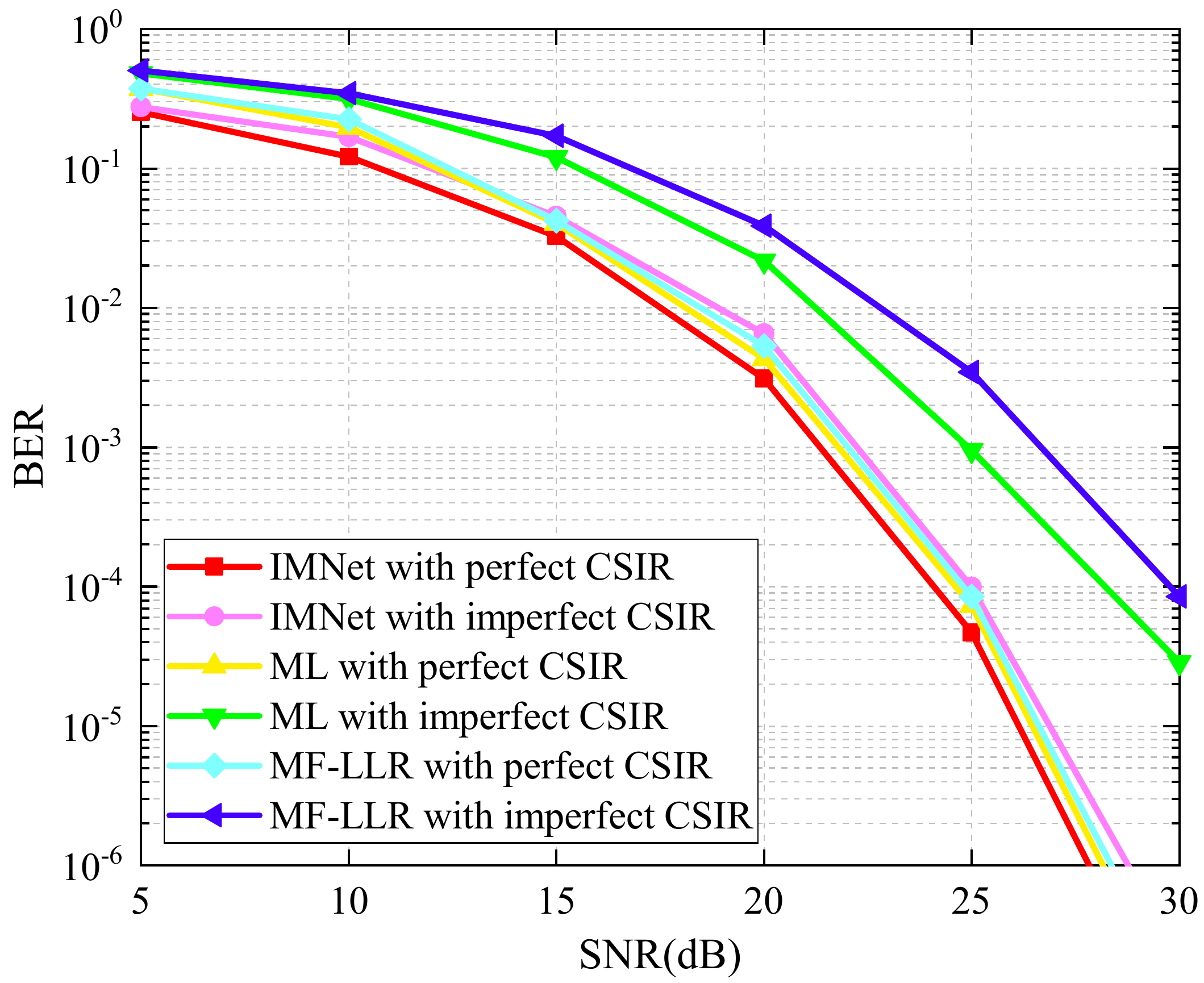}
	\caption{BER performance comparison of different approaches with correlated MIMO channel. IM-MIMO-OFDM is configured with $ N_t = N_r = 8, N_f = 8, K = 2, F = 6 $ and 4QAM modulation.}
	\label{fig-4.3}
\end{figure}



\subsection{Results}

\emph{1) Comparisons in Rayleigh fading MIMO channel:} Fig.\ref{fig-4.2} illustrates the BER performance of IMNet and two traditional algorithms (i.e., ML and MF-LLR) under Rayleigh fading MIMO channel. As expected, the proposed IMNet yields the best performance for both perfect and imperfect CSIR. When the CSIR is perfect, the BER performance of IMNet has about 2dB improvement compared with the traditional algorithms. When the CSIR varies from perfection to imperfection, the BER performance of ML and MF-LLR experience significantly decline, while IMNet achieves a more BER performance gain of over 3.3dB compared with the traditional algorithms at all SNR values. This gain comes from two aspects. One is that the AD subnet in IMNet can predict the activation probability of each TA without CSIR, which means the AD subnet is insensitive to channel variations. Therefore, when CSIR is imperfect, AD subnet is still able to provide accurate prediction of activation probability of each TA, which provides a solid foundation for the next step of signal detection. The other aspect is that the SD subnet can improve the accuracy of the recovered signal, which can further improve the BER performance.\par

\emph{2) Comparisons in correlated MIMO channel:} To further verify the adaptability and robustness of IMNet, we eavluate the BER performance of IMNet and two traditional algorithms (i.e., ML and MF-LLR) under correlated MIMO channel with perfect CSIR and imperfect CSIR. It can be seen from Fig.\ref{fig-4.3} that IMNet always achieves the best peroformance whether the CSIR is perfect or imperfect. When the CSIR changes from perfection to imperfection, there is a clear performance degradation greater than 10dB of traditional algorithms at 25dB SNR, while IMNet still maintains its BER at $ 10^{-4} $. The reason for is that traditional algorithms are much more dependent on the CSIR than the proposed IMNet, while IMNet can extract the characteristics of channel from the received signal and further improve the accuracy of the recovered signal.

\begin{table}[htbp]
	\caption{Parameters of three different IM-MIMO-OFDM scenarios.}
	\centering
	\normalsize
	\begin{tabular}{cccccc}
		\hline
		Scenario & $ N_t $ & $ N_r $ & $ N_f $ & $ K $ & $ F $ \\
		\hline
		\textbf{Scenario 1}  & 4 & 1 & 4 & 1 & 2 \\
		\textbf{Scenario 2}  & 8 & 2 & 8 & 2 & 6 \\
		\textbf{Scenario 3}  & 16 & 4 & 16 & 4 & 12 \\   
		\hline       
	\end{tabular}
	\label{table-1}
\end{table}

\begin{figure}[htbp]
	\centering
	\includegraphics[width=0.8\linewidth]{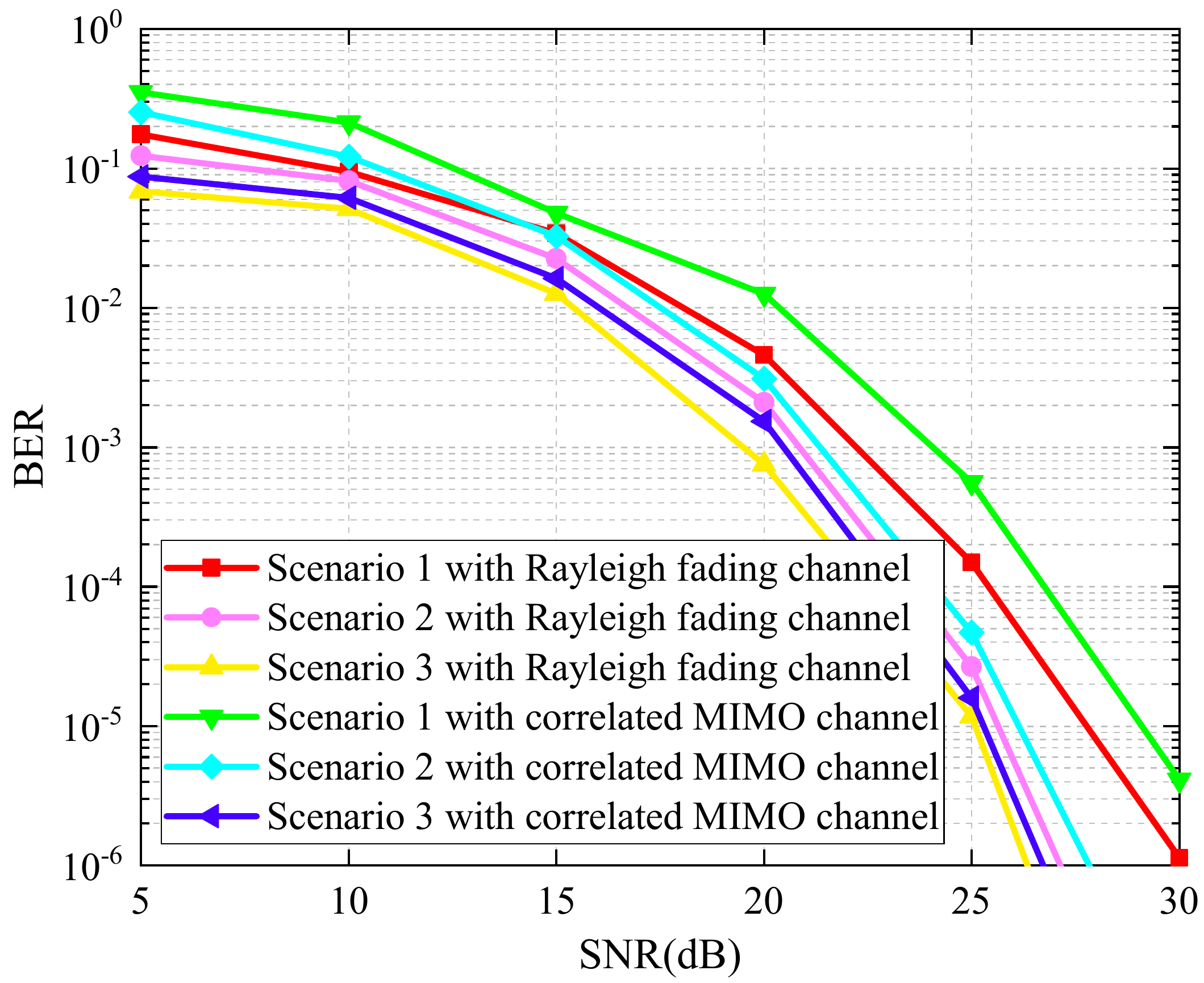}
	\caption{BER performance of IMNet in different IM-MIMO-OFDM scenarios. Modulation type is 4QAM.}
	\label{fig-4.1}
\end{figure}

\emph{3) Performance in different IM-MIMO-OFDM scenarios:} To verify the applicability of IMNet, we evaluate the BER performance of IMNet in three different IM-MIMO-OFDM scenarios under different SNR, and the results are shown in Fig.\ref{fig-4.1}. The configurations of three scenarios are detailed in TABLE \ref{table-1}, which make these scenarios become more complex gradually. We only consider perfect CSIR in each scenario. It can be recognized from Fig.\ref{fig-4.1} that IMNet has a fairly good performance in three scenarios. The BER is smaller than $ 10^{-3} $ for all scenarios at 25dB SNR. As the numbers of activated antennas and subcarriers increase, the BER performance has a certain gain by using proposed IMNet detector. Compared with Rayleigh fading channel model, the BER performance has a little degradation with correlated MIMO channel model in all three scenarios.\par

\emph{4) Comparisons in Computational Complexity:} To verify the low complexity of the proposed IMNet, the computational complexities of IMNet and two traditional algorithms (i.e., ML and MF-LLR) are compared in this part. We choose three different IM-MIMO-OFDM scenarios, shown in Table \ref{table-1} and the results are given in Table \ref{table-2}. We transmit 500 IM-MIMO-OFDM frames in each scenario. When the numbers of TA and subcarrier grows, we can activate more antennas and subcarriers to convey information, but the computational complexity will also be increased, which is consistent with the results in Table \ref{table-2}. We can also see that IMNet always achieves the lower computational complexity than the traditional algorithms in all three scenarios. In the first scenario where the transmitted information is the easiest to recover, MF-LLR needs about twice consumption of time than IMNet, and ML needs more than seven times consumption of time. As the scenario gets more complex (i.e., scenario 2 and scenario 3), the consumption of time of two traditional algorithms increases significantly, especially for ML algorithm. The reason for this phenomenon is that two traditional algorithms need a large number of iterations to search the optimum combination of the active TAs, the active subcarriers and the modulated symbols, while IMNet is a non-iterative detector that can directly predict the the active TAs and subcarriers and recover the signal.

\begin{table}[htbp]
	\caption{Computational complexity (in seconds) in different IM-MIMO-OFDM scenarios. The channel model is Rayleigh fading MIMO channel with perfect CSIR and SNR is 20dB.}
	\centering
	\normalsize
	\begin{tabular}{cccc}
		\hline
		Scenario & IMNet &  ML &  MF-LLR  \\
		\hline
		\textbf{Scenario 1}  & 7s & 52s  & 12s  \\
		\textbf{Scenario 2}  & 29s & 853s & 79s \\
		\textbf{Scenario 3}  & 127s & 2432 s & 1328s \\   
		\hline       
	\end{tabular}
	\label{table-2}
\end{table}

\section{Conclusion}\label{sec-5}
In this paper, we formulate the detection process of IM-MIMO-OFDM systems as a sparse reconstruction problem. Based on the structural sparsity of the transmitted signal and sparse reconstruction theory, a DL based non-iterative detector called IMNet is proposed to realize the detection process, while maintaining low complexity. The two most distinctive characteristics of IMNet are that its AD subnet can predict the activation probability of each TA without CSIR and its SD subnet can further mitigate the noise effects. These characteristics enable IMNet to achieve better performance with imperfect CSIR compared with the traditional algorithms. Besides, IMNet is a non-iterative detector, which make its detection complexity far less than the traditional algorithms. Simulation results demonstrate that IMNet outperforms two traditional algorithms in terms of BER and computational complexity in various scenarios, which verifies the better adaptability and robustness of IMNet.\par

\bibliographystyle{IEEEtran}
\bibliography{reference}

\end{document}